\documentclass[amsmath,amssymb,aps,pre,floatfix,twocolumn,superscriptaddress]{revtex4}

\usepackage{changepage}
\usepackage{mathtools}
\usepackage{floatrow}
\usepackage{graphicx}
\usepackage{dcolumn}
\usepackage{bm}
\usepackage{tikz}
\usetikzlibrary{shapes, arrows, calc, positioning,matrix}
\usepackage[export]{adjustbox}
\usepackage{bbold}
\usepackage{graphicx}
\usepackage{hyperref}
\usepackage[font=small,justification=RaggedRight,labelfont=bf]{caption}
\usepackage{enumitem}
\usepackage[utf8]{inputenc}
\usepackage[T1]{fontenc}
\usepackage{soul,xcolor}
\usepackage{float}
\usepackage[label font=bf,labelformat=simple]{subfig}
\usepackage{multirow}

\makeatletter
\newcommand{\subalign}[1]{%
  \vcenter{%
    \Let@ \restore@math@cr \default@tag
    \baselineskip\fontdimen10 \scriptfont\tw@
    \advance\baselineskip\fontdimen12 \scriptfont\tw@
    \lineskip\thr@@\fontdimen8 \scriptfont\thr@@
    \lineskiplimit\lineskip
    \ialign{\hfil$\m@th\scriptstyle##$&$\m@th\scriptstyle{}##$\crcr
      #1\crcr
    }%
  }
}
\makeatother

\makeatletter
\newcommand{\shorteq}{%
  \settowidth{\@tempdima}{-}
  \resizebox{\@tempdima}{\height}{=}%
}
\makeatother

\begin{document}

\title{Exogenous and Endogenous Price Jumps Belong to Different Dynamical Classes}

\author{Riccardo Marcaccioli}
\affiliation{Chair of Econophysics and Complex Systems, Ecole polytechnique, 91128 Palaiseau Cedex, France}
\affiliation{LadHyX UMR CNRS 7646, Ecole polytechnique, 91128 Palaiseau Cedex, France
}

\author{Jean-Philippe Bouchaud}
\affiliation{Chair of Econophysics and Complex Systems, Ecole polytechnique, 91128 Palaiseau Cedex, France}
\affiliation{Capital Fund Management, 23-25, Rue de l’Université 75007 Paris, France}
\affiliation{Académie des Sciences, Quai de Conti, 75006 Paris, France}

\author{Michael Benzaquen}
\email{michael.benzaquen@polytechnique.edu}
\affiliation{Chair of Econophysics and Complex Systems, Ecole polytechnique, 91128 Palaiseau Cedex, France}
\affiliation{LadHyX UMR CNRS 7646, Ecole polytechnique, 91128 Palaiseau Cedex, France
}
\affiliation{Capital Fund Management, 23-25, Rue de l’Université 75007 Paris, France}

\begin{abstract}
Synchronising a database of stock specific news with 5 years worth of order book data on 300 stocks, 
we show that abnormal price movements following news releases (exogenous) exhibit markedly different dynamical features from those arising spontaneously (endogenous). On average, large volatility fluctuations induced by exogenous events occur abruptly and are followed by a decaying power-law relaxation, while endogenous price jumps are characterized by progressively accelerating growth of volatility, also followed by a power-law relaxation, but slower than for exogenous jumps. 
Remarkably, our results are reminiscent of what is observed in different contexts, namely Amazon book sales and YouTube views. Finally, we show that fitting power-laws to {\it individual} volatility profiles allows one to classify large events into endogenous and exogenous dynamical classes, without relying on the news feed.
\end{abstract}

\maketitle

\section{Introduction}

Earthquakes, disease outbreaks, volcanic eruptions, avalanches, species extinctions, traffic jams, economic crises and financial crashes are but a few examples of a long list of extreme events that upend natural and social systems. Given their ubiquitous presence (and their relevance in our everyday life), they have received a great amount of attention from different scientific communities~\cite{bak_SOC_book,book_extreme_events,sornette_book_critical,taleb_book_black_sw}. A central question that researchers have tried to answer is whether these events are caused by  {\it exogenous} events (like the meteorite which probably triggered the Cretaceous–Paleogene extinction event) or result from some amplifying feedback mechanism internal to the system, in which case the shock is {\it endogenous}~\cite{sornette_chaper_endogenous}. 

This topic is particularly important in the context of financial markets, and is related to the long-standing Efficient Market controversy. If markets are efficient, significant price movements can only be due to unpredictable exogenous shocks. On the other hand, if self-reflexive feedback loops are present, extreme price displacements can be triggered by small (and seemingly irrelevant) fluctuations, which can ultimately generate substantial excess volatility. 

Is it really possible to categorize extreme events into exogenous and endogenous? Answering this question in a general context is highly non-trivial~\cite{sornette_chaper_endogenous}. Nevertheless, building on the idea that endogenous shocks can only appear in systems that are somehow ``fragile'', i.e. close to an instability, a methodology to differentiate exogenous from endogenous events in empirical data has been proposed in a very interesting series of papers~\cite{sornette2003endogenous}. 
Hawkes processes, in particular, provide a convenient and versatile modelling framework consistent with the assumption of a near-critical system. In fact, Hawkes processes were introduced to model self-exciting earthquakes \cite{hawkes_original,sornette_ETAS}, but have been shown to be relevant in many other contexts, such as financial market activity \cite{hawkes_mastromatteo}, crime outbursts \cite{mohler_hawkes}, or banking and corporate defaults \cite{lando_hawkes}, to name a few.

\begin{figure}
  \centering
  \includegraphics[width=0.9\linewidth]{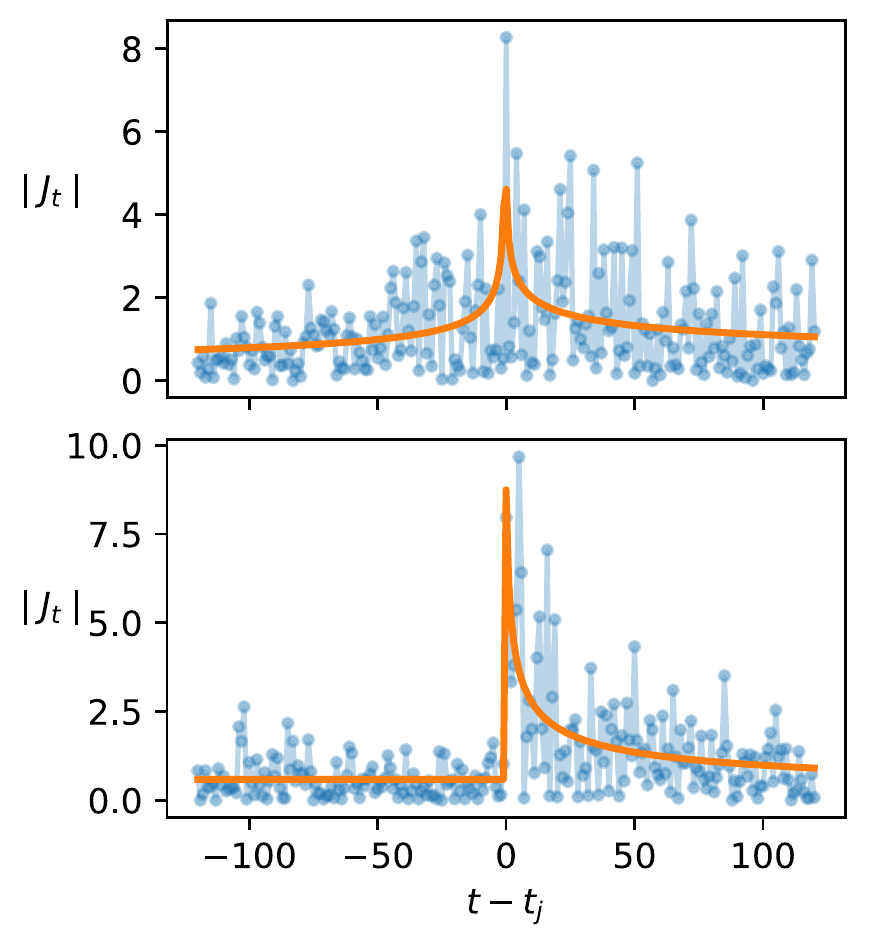}
  \caption{Examples of endogenous (top) and exogenous (down) bursts of volatility. We draw in orange the best fits using the functional form given by Eq.~\ref{eq: double pl} below. Exogenous shocks are characterised by a slow power-law precursory growth and an almost symmetric relaxation. Endogenous shocks are asymmetric around the instant of the shock and display a faster relaxation toward the pre-shock activity levels.}  
  \label{fig: sample endo exo noise}
\end{figure}

The proximity of an instability suggests the use of {\it critical} Hawkes processes, characterised by a power-law memory kernel. Such a specification has allowed the authors of~\cite{sornette_youtube} to efficiently discriminate endogenous from exogenous bursts of views among 5 millions YouTube videos. Views spikes which result from a contagion process on the underlying social network of influence are characterised by a slow (power-law) post-shock relaxation to the baseline views' number distribution and by an \emph{almost} mirror-image pre-shock growth. On the other hand, spikes that are chiefly induced by exogenous shocks are characterised by a faster post-shock decay, almost without any pre-shock growth (see Fig.~\ref{fig: sample endo exo noise} for an illustration in the case of market price jumps). Quite remarkably, similar results also hold for a dataset of Amazon books sales~\cite{sornette_books,amazon_book_sornette_long}. 

In the context of financial price time series, such alluring findings are still lacking. Nevertheless, several past studies hint at the possibility of dividing exogenous and endogenous extreme events in a similar manner. The observable of interest of most of these studies is the instantaneous volatility profile of stocks (log-)returns. First of all, it has been shown in~\cite{lillo_mantegna_pl,omori_law_1,omori_law_2,lillo_mantegna_2,pl_relax_2,zawadowski2006short} that the rate of abnormally large absolute returns after a large exogenous shock decays as a power-law (which corresponds to the so-called Omori law in the context of earthquake after-shocks~\cite{utsu_omori}). A similar behaviour has also been observed in the absolute value of the returns~\cite{lillo_mantegna_pl,vol_forecasting} (sometimes restricting to abnormally high returns~\cite{pl_relax_1,pl_relax_2,pl_time_asymmetry}) and in the dynamics of the bid-ask spread~\cite{pl_bid_ask,zawadowski2006short}. Quite strikingly, these findings seem to hold independently of the considered type of exogenous event, asset type or timescale over which returns are computed and therefore hint to the existence of a possible universality class. This is the ``Efficient Market Class'' (EMC), in the sense that the market price strongly reacts to unexpected large external shocks. Such strong reactions to exogenous events are not only limited to prices and are indeed known to be present in market activity as well~\cite{hisano2013high,lillo_news_hawkes_activity}.

Within the Efficient Market picture, all extreme events are exogenous, and no endogenous ``Self-Exciting Class'' (SEC) should be detectable. However, this view has long been challenged by a series of investigations, starting with the seminal work of Cutler, Poterba and Summers in 1989~\cite{vol_news_1} where they conclude that the {\it evidence that large market moves often occur on days without identifiable major news releases casts doubt on the view that stock price movements are fully explicable by news.} This conclusion, drawn using daily returns, was confirmed in later studies \cite{vol_news_2,vol_news_3} looking at different time granularity. In particular, the authors of Ref. \cite{vol_news_jp} found that most  large intra-day price fluctuations for single stocks (called ``jumps'' henceforth) happen independently from news releases. 

These latter findings point to the possibility that a non-efficient, SEC exists and that it can be successfully identified in empirical data, with jumps induced by a direct feedback loop between  trades and volatility. To expose the existence of such endogenous jumps, some studies take a granular approach and calibrate a Hawkes process on trade-by-trade data~\cite{filimonov2012quantifying,hardiman2013critical,wheatley2019endo,koyama2020statistical} while others focus on macroscopic quantities like volatility, volumes or trends~\cite{sornette2004volatility, quadratic_hawkes, fosset2020endogenous}. 

In this study we follow this latter stream of literature. In particular the present work builds upon the observations presented in Ref.~\cite{vol_news_jp}. By using a database of news concerning single or multiple stocks, it has been empirically established that the average volatility profile following a news-induced, EMC jump is markedly different from the one following an SEC jump.

We broadly confirm and significantly extend the results of Ref.~\cite{vol_news_jp}, using a more recent database (300 different stocks traded at the NYSE from 01/01/2015 to 01/01/2020). We show that it is indeed possible to classify extreme price moves into two distinct dynamical classes, EMC and SEC. Inspired by seismology studies, we argue that instead of individual price jumps and individual news one needs to focus on clusters of jumps and clusters of news. We show how SEC clusters of jumps (not triggered by news) display very different properties from EMC clusters (closely following a cluster of news). Consistent with previous studies on YouTube views and Amazon book sales \cite{sornette_youtube, sornette_books}, we find that the average volatility profile of SEC events is much more symmetric than the average profile of EMC events and that, while they both decay as power-laws, they are characterised by different relaxation exponents.
In fact, the observed values of these exponents are remarkably close to those reported for YouTube views and Amazon book sales \cite{sornette_youtube, sornette_books}. In agreement with the recently introduced endogenous liquidity crises model~\cite{fosset2020endogenous}, SEC cluster of jumps appear to be preceded by a slow increase in volatility and price trends. 

Having established the {\it average} profile of EMC and SEC jumps, we then turn to analyzing \emph{individual} volatility profiles around large clusters of jumps. Determining the shape (parameterized by a power-law before and after the first jump of a cluster) and asymmetry of these profiles allows us to classify jumps into EMC and SEC types with remarkably high degree of success, as measured by the Area Under Curve of the corresponding classification tasks. Finally, we discuss several wider implications of our findings.

\section{Extreme events in electronic markets}

\subsection{The Limit Order Book}

In the present days, most of the world’s financial markets use an electronic trading mechanism called a Limit Order Book (LOB) to facilitate trade of a given asset. 
The LOB $\mathcal{L}(t)$ is the collection of all active limit orders at any given time $t$ and, as such, it can be thought as a concise representation of the supply and demand of any electronically tradable asset. The LOB is usually divided into the ask side (the set of active sell limit orders) and the bid side (the set of active buy limit orders). The highest (lowest) occupied buy (sell) price level is called the ask $a_t$ (bid $b_t$). The difference between the two is called the spread $s_t = a_t-b_t$ while their average $m_t = (a_t+b_t)/2$ is called the mid-price and it is usually used as a proxy for the price of an asset (see \cite{micro_price} for more on this topic). 
Each price level at or below the bid (above the ask) is populated by volumes $V_b,V_{b-1}, \dots V_{b-n}, \dots$ ($V_a,V_{a+1}, \dots V_{a+n}, \dots$). This means in particular that a buy market order of size $Q_n=V_a+V_{a+1}+ \dots+ V_{a+n}$ leads to an immediate ask price move up by $n$ ticks. (Note that some price levels maybe empty). The quantity $n/Q_n$ can be thought of as a measure of sparsity of the LOB on the ask side, with a similar definition for the bid side. 
We refer the interested reader to Ref.~\cite{lob_review_1, lob_review_2, bouchaud2018trades} for extensive reviews on the empirical properties of LOBs.

\subsection{Dataset description}

\subsubsection{Order book data} 

We conduct the analysis by using the four best price levels (2 for the bid and 2 for the ask) from the LOB of a selection of 300 stocks continuously traded on the NYSE from 01/01/2015 to the 01/01/2020. Each LOB is sampled on a minute timescale. These snapshots portray the time evolution of the price and the supply and demand of a given stock. We only consider data collected during the regular US trading session (which start at the 9:30 a.m. and ends at the 4:00 p.m.) and only those sessions with a moderate or high trading activity (we only keep trading days with at least 300 recorded price changes). The reasons for this latter filtering step are threefold. First of all, to uniformise our sample: some stocks are always characterised by a continuous moderate or high trading activity while others are not. Secondly, to avoid spurious effects: our jump detection methodology assumes that the returns (after standardization) are approximately distributed as a standard normal; this assumption crumbles when the market activity is low, leading to the detection of numerous spurious jumps~\cite{lillo_cojump}. Lastly, estimation accuracy: estimating scaling laws is a notoriously hard task~\cite{sornette_youtube}, especially in highly noisy environments~\cite{horbelt_scaling}. We noted that removing from our samples those days with an exceedingly high amount of zeros or missing values in the volatility series led to a more accurate estimation process. Coherently with the moderate activity requirement, we selected the stocks based on their 2019 turnover. For a complete list of all the stocks included in our analysis, as well as a detailed description of the data pre-processing, see the Appendix.

\subsubsection{News data} 

We use a generic (i.e. not only finance related) news database which contains articles published on Bloomberg during the same period (01/01/2015 to 01/01/2020). Each news item is characterised by its title, the time at which it was been posted online and a list of tickers (i.e. unique stocks identifiers) which the news may concern. For this study we will only consider those news which are marked as relevant for at least one of the 300 stocks we consider and which explicitly display in the title the ticker of 
\begin{enumerate}
    \item at least one of the stocks it may concern, or
    \item at least one of their companies' names, or
    \item at least one of their companies' abbreviated names (i.e. IBM instead of International Business Machines, or Abbott instead of Abbott Laboratories).
\end{enumerate} 
See the Appendix for summary statistics of the news, as well as their distribution across times and stocks.

\subsection{Price Jumps Detection}

The observable we shall focus on in our analysis is the mid-price $m_t$. Before exposing possible differences between exogenous and endogenous extreme mid-price movements, we need a way to assess which variations $m_{t}-m_{t-1}$ can be considered extreme or abnormal. In order to do so, we follow the non-parametric price jumps detection methodology proposed in Ref.~\cite{jump_test_1} and further refined in Ref.~\cite{jump_test_2}. The intuition behind such procedure is very straightforward: fluctuations of the mid-price $m_t$ are first normalized so that, in the absence of jumps, their distribution is as close as possible to a standard normal distribution. Once this normalization is properly defined, Extreme Value Theory can be used to derive a threshold above which a fluctuation can be classified as a jump within a given probability level. 

We consider the 1-minute return time series $r_t = \log{\frac{m_t}{m_{t-1}}}$. Mid-price returns are known to have approximately zero mean but a strongly fluctuating variance, with both intra-day seasonalities and long-memory, intermittent dynamics (see e.g. \cite{cont_empirical, MRW, rough_vol_long_range, hardiman2013critical, quadratic_hawkes}). As such, any standardization procedure must take into consideration both the instantaneous evolution of the variance as well as any possible seasonality. We therefore define the ``jump-scores'' $J$ as:
\begin{equation}\label{eq: J stat}
    J_t = \frac{r_t}{\sigma_t f_t} \; ,
\end{equation}
where $\sigma^2_t = \frac{\pi}{2 K} \sum_{i=1}^{K} \vert r_{t-i} \vert\, \vert r_{t-i+1}\vert$ is an estimator of the local volatility over a rolling time window of length $K=390$ (i.e. one day worth of data, but dropping any overnight contribution) and $f_t$ is an estimator of the intraday periodicity component (see the Appendix for its detailed definition). 

Under the null hypothesis of no jumps and a vanishing sampling frequency, the statistics of the maximum of $\vert J_t \vert$ converges to a Gumbel distribution. One can therefore reject, with a statistical significance $\alpha=0.01$, the null hypothesis of absence of jumps whenever we observe:
\begin{equation*}
\vert J_t \vert\, > C_{K} - S_{K} \log{ (\log{\frac{1}{1 - \alpha}} ) }
\approx 4.36, \quad (K=390).
\end{equation*}
The constants $S_K =(2 \log K )^{-0.5}$ and $C_K = (2 \log K )^{0.5} - (\log \pi + \log(\log K))/(2 (2 \log K )^{0.5}) $ are dependent on the window size and are meant to correct for the fact that, within each window, we are performing multiple hypothesis tests. As such, by using this threshold, one expects to find only $\alpha$ spurious jumps in a given sample of $K$ observations.

In a nutshell, we mark as ``jumps'' those price movements with associated z-scores that are approximately 4-sigma away from zero. In order to avoid effects due to market opening or closing, we discard jumps happening in the first or last 15 minutes of the trading day.

\subsection{Clusters of jumps}

\begin{figure*}[t]
  \centering
  \includegraphics[width=0.85\linewidth]{fig1.pdf}
  \caption{Characterization of identified clusters of jumps. \textbf{(a)} Inter-times distribution of jumps and clusters of jumps. We normalize with the minimum time granularity to detect the given event, i.e one minute for jumps and $L$ minutes (Eq.~\ref{eq: cluster limit}) for clusters of jumps. The blue line is the best fitting power law (exponent 0.88) and the red line is the best fitting exponential (rate 0.11). \textbf{(b)} Distribution of the number of jumps within exogenous (\textquotedblleft news\textquotedblright) and endogenous (\textquotedblleft no news\textquotedblright) clusters. \textbf{(c)} Distributions of Kendall's tau rank correlations between the amplitude ranking of the jumps in a given cluster with $N$ jumps and their  chronological ranking. }  
  \label{fig: cluster 1}
\end{figure*}

We run the price jumps detection methodology outlined above on all the 300 mid-prices time series. We record a total of 258,671 jumps. The daily average number of jumps of a given stock ranges from $0.25$ to $3.26$, with an average value of $0.70$ and a standard deviation of $0.42$.  

As such, on average, we would expect a stock to jump about once per day (which is, in passing, much more frequent that the expected number of news that can shake the value of a given stock). However, if we look at the inter-time distribution between two consecutive jumps within the same day (Fig.~\ref{fig: cluster 1}a)) we observe clear deviations from a Poisson law. Rather, the distribution is well fitted by a power-law behaviour. Power law distribution of waiting times is a typical fingerprint of many social activities~\cite{barabasi_burstiness}, including trading in financial markets, but also seismic activity or epileptic activity, see e.g.~\cite{saichev,chicheportiche}. Such ``bursty'' time series are often modelled in terms of self-exciting Hawkes-like point processes~\cite{hawkes_mastromatteo}. Indeed, our empirical findings are consistent with previous observations, see e.g.~\cite{lillo_cojump}. 

Such long-memory effects in the dynamics of jumps can potentially induce spurious effects when an aggregate analysis is performed. Consequently, and following common practice in seismology~\cite{sornette_ETAS}, we move away from earlier studies on price jumps~\cite{boudt2014intraday,vol_news_jp} and instead of analysing single jumps, we focus on {\it clusters} of jumps. 

We adopt a simple and intuitive clustering technique to group jumps together: we compare the observed inter-times between any two consecutive jumps against the one prescribed by a Bernouilli null-hypothesis, corresponding to independent jumps occurring with probability $p$. If, under the null hypothesis, the probability of observing the given inter-time is smaller than a significance level $\epsilon$, we cluster the two jumps together. 

It is straightforward to show that, under this simple null model, given the presence of a jump at time $t_0$, a second jump occurring at time $t_1$ is assigned to the same cluster when:
\begin{equation}\label{eq: cluster limit}
t_1-t_0 <  \frac{\log \left( 1-\epsilon \right)}{\log \left( 1-p \right)} - 1 \, .
\end{equation}
We set $\epsilon=0.05$ and we determine $p$ in order to have our null model preserving, on average, the number of jumps of each stock within any given month.

After running our clustering methodology, we find a total of 197,197 clusters of jumps (most made out of one single jump). The daily average number of clusters of jumps for a given stock ranges from $0.17$ to $2.77$ with an average value of $0.53$ and a standard deviation of $0.37$. The normalized inter-time distribution of those clusters happening in the same day is now well described by an exponential distribution (Fig.~\ref{fig: cluster 1}a). This feature validates that such clusters can be reasonably considered to be independent, and therefore that spurious effects induced by any aggregation procedure are reasonably reduced.

\subsection{News Related Jumps} 

Similarly to jumps, we also observe that news releases tend to cluster in time. We therefore perform on news the same clustering procedure applied to jumps. We then mark as news related (or exogenous) those clusters of jumps which started up to one minute before and up to four minutes after the beginning of a cluster of news. This is done in order to account for the fact that a particular news may have become available to some market participants before our recorded news release timestamp and to account for possible misalignments between the news feed and the LOB data. We mark as not news-related, or endogenous, the remaining clusters of jumps. For simplicity, we exclude from our analysis those endogenous clusters which start within a cluster of news. 

Another effect that one should need to consider is the role of macroeconomic news (not specific to a given stock), which might trigger clusters of stock price jumps. A systematic identification of possibly relevant macroeconomic news would entail contextual word recognition and goes beyond the scope of the current work. To circumvent such a limitation, we remove from our list of clusters those which participate in a market-wide or sector-wide event, as any relevant macroeconomic news would trigger. Hence we compute, for each cluster of jumps, the number of stocks which display an overlapping cluster of jumps during the same time interval. Whenever we observe a number of overlapping clusters higher than 30 (10\% of the stocks in our pool), we mark them as belonging to a market-wide or sector-wide event. Changing this threshold to 15, 60 or to one prescribed by a null hypothesis of clusters independence (detailed in the Appendix), does not significantly affect our findings. 

As a final safe-guard, we also exclude from our samples  clusters of jumps, of a given stock, happening within 100 minutes from each other. This is done following Ref.~\cite{boudt2014intraday} in order to completely avoid any contamination effect that may happen in our analysis.  

Finally, we are left with a total of 106,680 clusters of jumps, out of which only 1073 are news related (note that most major, company related news happen outside market hours).

\section{Results}

\subsection{The Internal Structure of Clusters}

We now compare the composition of endogenous and exogenous clusters. In Fig.~\ref{fig: cluster 1}b, we observe that news-related clusters are characterised by a higher average number of jumps. In Panel Fig.~\ref{fig: cluster 1}c, we compare the distribution of Kendall's tau correlation~\cite{kendall1938new} between the chronological ranking of jumps and the ranking of jumps based on their amplitude. A value $\tau=1$ corresponds to the case where the jump happening first is also the largest, the second jump is the second highest and so on. A value $\tau=-1$ corresponds to a sequence of jumps happening in ``reverse order'', the largest one being the last of the cluster. 
Beside accommodating more jumps, news-related clusters are more naturally ordered in time than endogenous clusters, consistent with the idea that exogenous events are strong and sudden responses to an external shock while endogenous events are the result of a self-exciting stochastic process, with a progressive build-up. 

Motivated by this consideration, and by the well-documented assertion that instantaneous mid-price variations can be described by means of Hawkes processes (see Ref.~\cite{hawkes_mastromatteo} for a recent review), we now show how exogenous (EMC) and endogenous (SEC) events are characterised by markedly different profiles of instantaneous volatility, price trend and LOB sparsity. 

\subsection{Average Profile of EMC and SEC Jumps}

\begin{figure*}
  \centering
  \includegraphics[width=0.9\linewidth]{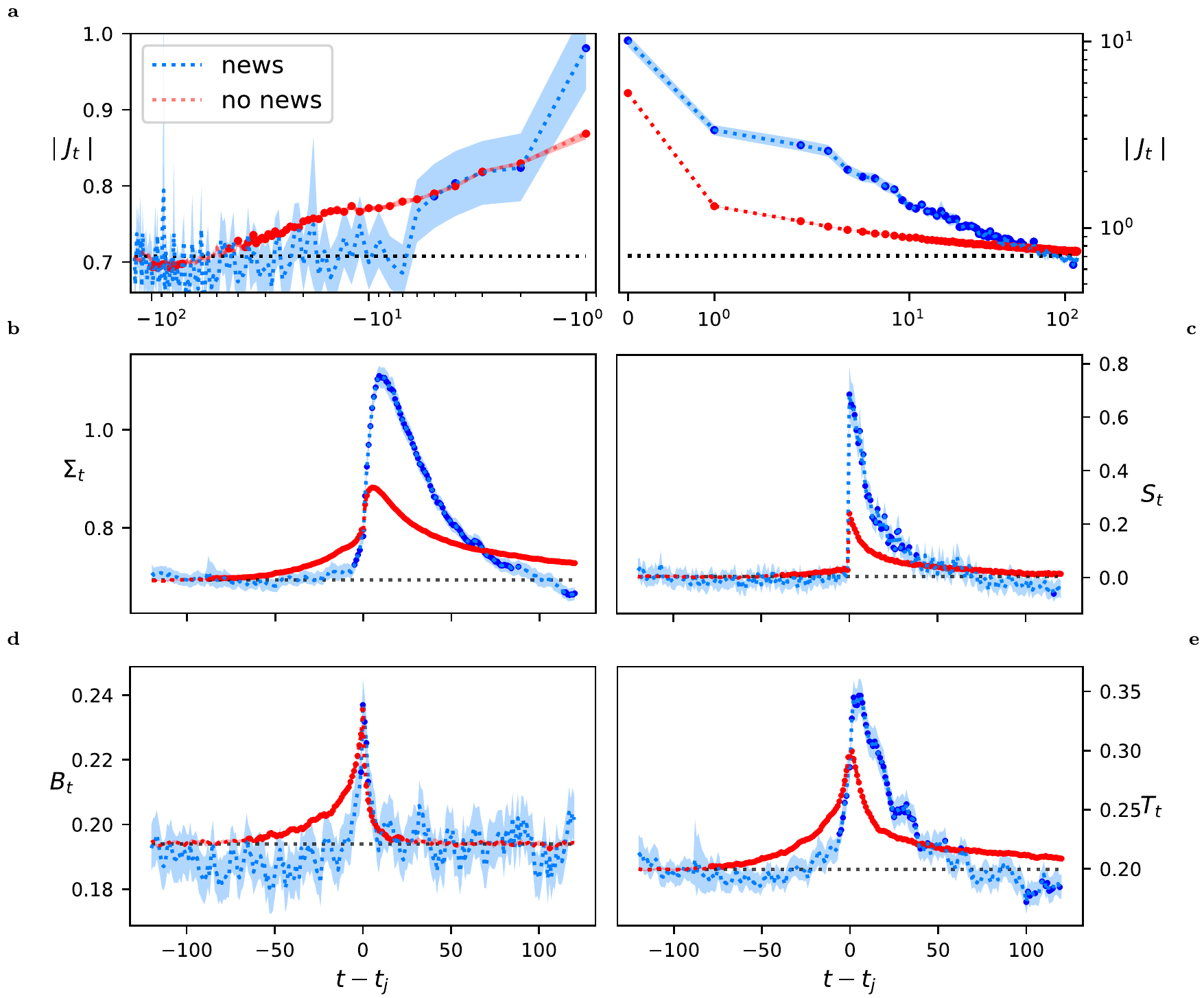}
  \caption{Differences between clusters of jumps which happened in close proximity to a news release and those which did not. \textbf{(a)} In blue we show the average of the absolute jump-score $\vert J_t \vert$ only when news related clusters are considered, in red we highlight the average across the remaining clusters. The $0.01$ confidence bands (lighter colours) on those averages are obtained using bootstrapped samples. For each point in time and for each of the two sub-samples, we perform a Welch's $t$-test against the distribution of $\vert J_t \vert$ coming from the first 20 minutes of our observation window. We use a marked dot whenever we can reject, at $0.01$ significance, the null hypothesis that the two distributions have the same mean. We apply the FDR method to account for the multiple tests performed. Note that the power of the test is different between the news and no news case given their different sample sizes. We use a log scale to highlight the power law behaviour of the instantaneous volatility. \textbf{(b)} Same plot for past excess volatility $\Sigma_t$, in linear scale. \textbf{(c)} Same plot for the instantaneous normalized LOB sparsity $S_t$, in linear scale. Notice the small decrease of liquidity starting 15 minutes before no-news jumps. \textbf{(d)} Same plot for past binarised trends $B_t$, in linear scale. \textbf{(e)} Same plot for past trends $T_t$, in linear scale. Note that the averaging is performed on the absolute values of $T_t$ and $B_t$. }
  \label{fig: aggregated}
\end{figure*}

In order to show that clusters of jumps which are triggered by a news release display markedly different characteristics from those which are not, we focus on the following five quantities: 
\begin{itemize}
    \item Instantaneous jump-score: $\vert J_t \vert$;
    \item Exponential moving average of past excess volatility, defined as:
    \[
    \Sigma_t = \kappa \vert J_t \vert+ (1-\kappa) \Sigma_{t-1} \,,
    \]
    where $\kappa$ defines the averaging timescale, here chosen to be $\kappa=0.12$ (corresponding to a decay time of 16 minutes, see~\cite{fosset2020endogenous,quadratic_hawkes}). Note that we exclude from the exponential  average calculations the standardized returns $J_t$ marked as a jumps;
    \item Normalized past price trend:
    \[
    T_t = \kappa J_t + (1-\kappa) T_{t-1} \,,
    \]
    using the same value of $\kappa$ as above, and the same exclusion of jumps in the computation;
    \item Binarized past price trend:
    \[
    B_t = \kappa \frac{J_t}{\vert J_t \vert} + (1-\kappa) B_{t-1} \,,
    \]
    using the same value of $\kappa$ as above;
    \item Instantaneous average LOB sparsity $s_t$, defined using the 2 best limit prices:
    \[
    s_t = \max\left[ \frac{p_t^{a} - p_t^{b+1}}{\psi (1+\log V_t^b)} \, , \, \frac{p_t^{a+1} - p_t^{b}}{\psi (1+\log V_t^a)}  \right] \,,
    \]
    where $\psi$ is the tick size (here $\psi=0.01 \$$ for all stocks in our sample). Note that we define the sparsity using the less dense side of the LOB. Given that $s_t$ is sensitive to the local market activity and posses non-negligible intra-day periodicity, we define the associated z-score 
    \[ 
    S_t = \frac{s_t}{f_t \sigma_t} - \mu_t,
    \] 
    where $\sigma_t$ is the standard deviation of $s_t$ over a rolling window including the last day worth of data ($K=390$), $f_t$ is the average value of $s_t/\sigma_t$ across all the points with the same intra-day periodicity and $\mu_t$ is the average value of $s_t / f_t \sigma_t$ over a rolling window including the last day worth of data
\end{itemize}

In Fig.~\ref{fig: aggregated}, we show how the average profiles of these five measures differs when calculated only on SEC clusters of jumps (not preceded by any news) or only on EMC clusters (in close proximity of a news). Averaging is done by shifting time such that for each cluster, $t=0$ corresponds to the {\it first} jump of the cluster. Given that we are not interested in the sign of past trends, but only their magnitude, the averaging is performed on the absolute values of $T_t$ and $B_t$.

The panels clearly show that SEC clusters are preceded by a slow  increase of volatility and trends. The volatility increase starts to be statistically relevant up to 75 minutes before the occurrence of the first jump. The sparsity of the order book does increase, albeit weakly, 15 minutes before no-news jumps (see Fig.~\ref{fig: aggregated}, panel c). We however expect that the final drop of liquidity takes place at higher frequency, due to the fierce competition between High-Frequency liquidity providers. 

EMC clusters, on the other hand, happen much more abruptly, and their average profiles before the first jump hardly show any increase at all for all five metrics. 

Consistent with observations for other social systems, we also see a clear difference in the relaxation of the volatility {\it after} the first jump. Endogenous clusters, even containing fewer jumps, revert to the average baseline volatility more slowly than exogenous clusters. Relaxation after EMC jumps is not only faster, it actually undershoots the baseline volatility: two hours after the first jump, four out of five indicators appear to be lower than the values recorded two hours before the first jump. This was also noted in Ref.~\cite{vol_news_jp}, and interpreted by arguing that after the release of news, uncertainty about price is actually reduced. In contrast, endogenous jumps cannot be rationalized by market participants, and uncertainty remains high for a longer period.

Following previous work \cite{sornette_youtube, sornette_book_critical, lillo_mantegna_pl,zawadowski2006short,pl_relax_2, vol_news_jp} we now quantify the speed of the pre-jump and post-jump dynamical profiles by fitting a double power-law function of the form:
\begin{equation}\label{eq: double pl}
\vert J_t\vert= f(t) =
\begin{cases}
&\frac{N_\ell}{\vert t-t_{\text{c}} \vert^{p_\ell}} + d, \qquad (t < t_{\text{c}}) 
\\
&\frac{N_r}{\vert t-t_{\text{c}} \vert^{p_r}} + d, \qquad (t > t_{\text{c}}),
\end{cases}
\end{equation}
where $d$ is the baseline volatility, $t_{\text{c}} \in [t_{\text{j}}-1,t_{\text{j}}]$ is the time of the shock and $t_{\text{j}}$ is the time at which the first jump of the cluster occurs. To estimate the coefficients appearing in Eq.~\eqref{eq: double pl}, we use a non-linear least squares fitting and discard the first jump of a cluster. We then check that the normalized residuals are normally distributed using a Shapiro-Wilk test~\cite{shapiro1965analysis} at the $0.01$ statistical significance.

For the exponents, we find $p_\ell=0.36 \pm 0.02$ and $p_r=0.40 \pm 0.02$ for SEC clusters and $p_\ell=0.08 \pm 0.01$ and $p_r=0.68 \pm 0.01$ for the EMC clusters. These values of $p_r$ are different, but not very far, from those reported in \cite{vol_news_jp}, i.e. $p_r \approx 0.5$ for SEC and $\approx 1$ for EMC.  

For the amplitudes $N_{\ell/r}$, we find $N_\ell = 0.235 \pm 0.009$ and  $N_r = 0.59 \pm 0.01$ for SEC clusters and $N_\ell = 0.68 \pm 0.01$ and  $N_r = 4.76 \pm 0.06$ for EMC clusters. The estimated SEC baseline volatility is $d = 0.66 \pm 0.01$ while we find $d = 0.48 \pm 0.01$ for EMC. This discrepancy in the baseline volatility is because, as we can observe in Figure~\ref{fig: aggregated}, the asymptotic post-cluster volatility of the EMC cluster is lower than the pre-shock baseline.\footnote{Leaving $d_r$ and $d_\ell$ free does not significantly change the values of the exponents $p_r$ and $p_\ell$.} The estimated jump time of the SEC cluster $t_{\text{c}} = t_{\text{j}} \pm 0.08$ coincides with the time of the first jump of a cluster while, for the EMC class, we find $t_{\text{c}} = t_{\text{j}}-1.00 \pm 0.03$, which is consistent with a pre-shock explosive growth.

Note that the relaxation of the LOB sparsity after a jump can also be fitted by a power-law with $p_r^{S} \approx 0.4$ for SEC jumps and $p_r^{S} \approx 0.7$ for EMC jumps, not very different from the ones governing the relaxation of $|J_t|$. On the other hand, the pre-jump SEC profiles start picking up too close to $t_{\text{j}}$ to allow for a meaningful fit with our one minute resolution time.

\subsection{Predictions of a Hawkes Model}

Besides confirming the different volatility relaxation speeds between endogenous and exogenous jumps, as in other studied systems, our results show that there exist an asymmetry between the pre-jump growth and the post-jump relaxation {\it even for SEC}. To rationalize their findings Refs.~\cite{sornette_youtube,sornette_books}  postulate the existence of a  self-exciting process of the form
\begin{equation}\label{eq: hawkes simple}
\begin{aligned}
\lambda(t) =  \lambda_0(t) + \sum_{t_i < t} \phi(t-t_i)  \; ,
\end{aligned}
\end{equation}
where $\lambda(t)$ the instantaneous rate of price moves, $\lambda_0(t)$ is the exogenous rate of price moves, $t_i$ is the time at which previous price moves took place, and $\phi(\tau)$ is the memory kernel of the system, which captures the way past events enhance the probability of current events. 

The rationale for using the self exiting process of Eq.~\eqref{eq: hawkes simple} is fairly intuitive. Given the fact that the LOB is a public source of information, any change (exogenous or endogenous) in the instantaneous volatility may trigger a reaction in some market participants whose actions will have an effect on the volatility itself which will then trigger a second generation effect on other market participants and so on and so forth. This impact of a trader onto other traders, or of past volatility onto future volatility, is not instantaneous and it is modelled by the memory kernel $\phi(t-t_i)$.

By simply assuming a power-law memory function $\phi(\tau) \sim 1/\tau^{1+\theta}$ with $0<\theta<1$, Eq.~\eqref{eq: hawkes simple} elegantly predicts two different profiles for exogenous and endogenous jumps~\cite{helmstetter2002subcritical,sornette_books} when the process is marginally stable (i.e. when $n := \int_0^\infty {\rm d}\tau \phi(\tau) \to 1$). One finds the following behaviour for the pre- and post-jump profiles:
\begin{equation}\label{eq: hawkes pred}
   \vert J_t \vert \propto 
   \begin{cases}
   (t - t_{\text{j}})^{\theta - 1}, \quad &\text{EMC}, \quad t > t_{\text{j}}, t - t_{\text{j}} \ll (1-n)^{-\frac{1}{\theta}};\\
   (t - t_{\text{j}})^{-\theta - 1}, \quad &\text{EMC}, \quad t > t_{\text{j}}, t - t_{\text{j}} \gg (1-n)^{-\frac{1}{\theta}};\\
   \vert t - t_{\text{j}}\vert^{2\theta - 1}, \quad &\text{SEC}, \quad t \lessgtr t_{\text{j}},
   \end{cases}
\end{equation}
with a flat profile (no precursor) for EMC, $t < t_{\text{j}}$.

Comparing these predicted profiles with the average ones plotted in Fig.~\ref{fig: aggregated}, we see that a post-jump dynamics with $\theta=0.3$ is consistent with our data for which $p_r^{SEC} \approx 0.4 = 1-2\theta$, and $p_r^{EMC}\approx 0.7 = 1-\theta$, positing, as argued in \cite{hardiman2013critical}, that financial markets are indeed close to criticality ($n \approx 1$). Note that Wehrli et al. \cite{Wehrli2021} have recently questioned this assumption, asserting that the low frequency kernel contribution to $n := \int_0^\infty {\rm d}\tau \phi(\tau)$ is dominated by the exogenous dynamics of the rate $\lambda_0(t)$ in Eq. \eqref{eq: hawkes simple}. While this may well be the case, we satisfy ourselves in this work with the idea that critical Hawkes processes provide an {\it effective} description of feedback effects in financial markets, and treat Eq. \eqref{eq: hawkes pred} as a convenient fitting function. 

Within this framework, the value we observe for $\theta$ is, quite remarkably, exactly the same as the one reported in previous studies on other social systems~\cite{sornette_youtube,sornette_books}. Note that our value for the relaxation exponent $p_r^{SEC} \approx 0.4$ is quite close to the one estimated in Ref.~\cite{zawadowski2006short}, where post-jump volatility profiles of liquid US stocks were also studied (in the period 2000-2002). 

However, the pre/post jump symmetry predicted by the model for SEC jumps is (mildly) violated -- see the values of $N_\ell$ and $N_r$. We conjecture that such an asymmetry could be captured by the generalization of Eq.~\eqref{eq: hawkes simple} recently proposed in~\cite{quadratic_hawkes}, where not only past activity, but also past price trends, feedback on the current rate of activity. Such a coupling indeed leads to a measurable time reversal asymmetry in the volatility dynamics~\cite{quadratic_hawkes} and therefore can potentially produce an asymmetry between pre- and post-shock volatility dynamics as the one we observe \footnote{Note that, as measured in Refs.~\cite{quadratic_hawkes,fosset2020endogenous} the strength of the feedback between past trends and future volatility is much smaller than the one between past and future volatility. This is compatible with the observed weak degree of asymmetry between pre- and post-jump profiles.}. We leave this question open for further investigation. It is worth mentioning that another possible explanation for such asymmetry is that a non negligible portion of the jumps we marked as endogenous are driven by exogenous information not detected by our news database. 

\begin{figure*}
  \centering
  \includegraphics[width=0.8\linewidth]{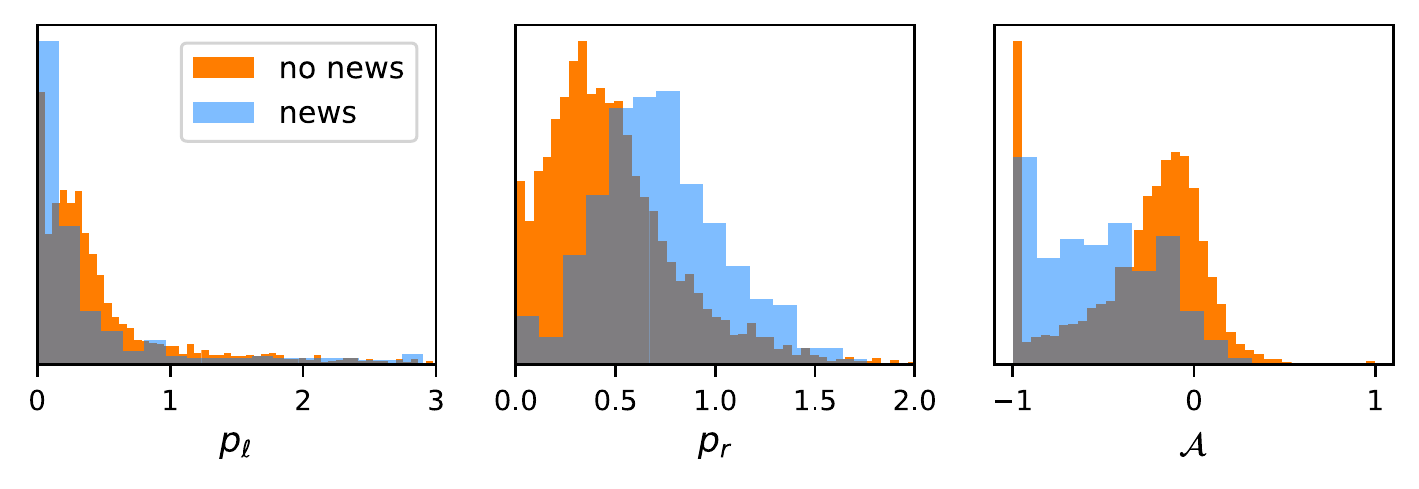}
  \caption{Results of the double power-law fitting on the instantaneous volatility profiles. The name of the best fitting parameter each plots represents can be read directly on the plots.}  
  \label{fig: fit hists}
\end{figure*}

\section{Classification of Single Volatility Profiles}

In this final section we show that, even in a highly noisy environment such as financial markets, the classification of different jumps into SEC and EMC provided by the news feed can be successfully reconstructed only using individual volatility profiles.

Even if the Hawkes model (Eq.~\eqref{eq: hawkes simple}) is not fully compatible with average profiles, as shown in the previous section, we attempt to use the functional forms suggested by Eq.~\eqref{eq: double pl} to fit {\it individual} volatility profiles and infer from such fits the nature of the observed events. 

To do so, we fit the functional form of Equation~\ref{eq: double pl} to each single volatility profile. In line with Refs.~\cite{sornette_books,sornette_youtube} and with the discussion of the previous section, we expect the following characteristics:
\begin{itemize}
\item $p_r^{EMC} \approx 1-\theta > p_r^{SEC} \approx 1 - 2\theta$:   higher relaxation exponents for those clusters which happen in close proximity of the release of a piece of news.
\item $p_\ell^{EMC} \ll 1$ or $\gg 1$: the unanticipated, explosive nature of EMC jumps leads to a numerically determined exponent that is either very large or very small. 
\item Defining the asymmetry $\mathcal{A}$ of a jump from integrated area under the pre- and post- region of the profiles (see the Appendix for an operational definition) we expect $\vert \mathcal{A}^{EMC} \vert > \vert \mathcal{A}^{SEC} \vert$: the endogenous class is characterised by a rather symmetric pre-jump growth and post-jump relaxation, while exogenous events are strongly asymmetric.  
\end{itemize}

Naturally, we do not expect such a sharp distinction between EMC and SEC at the level of individual events. For example, there are cases where the news leaks before announcement, leading to an increase of volatility ahead of the jump. Conversely, some endogenous events may show very little pre-jump activity since they can be triggered by ``fat-fingers'', by rogue algorithms or by some exogenous piece of information not present in our news database. Nevertheless, we will show that a relatively robust classification can still be performed by only considering the shapes of the single volatility profiles.

Given the high level of noise in price movements, we restrict our analysis to relatively high intensity clusters, i.e. to those clusters made up of at least two jumps. This leave us with a total of 10,491 clusters of jumps, out of which 391 happened in proximity of news releases. For each volatility profile, we perform a direct non-linear least squares fit of Eq.~\ref{eq: double pl} (see the Appendix for a detailed description of the procedure) and we keep only those fits with a median relative error on the coefficients smaller than one. This leaves us with 5,461 SEC and 321 EMC events. In Figure~\ref{fig: fit hists}, we plot the empirical distribution of the fitted values of $p_{\ell}$, $p_r$ and $\mathcal{A}$ for both types of events.

First of all, we observe that the results we obtain are remarkably consistent with the results on average profiles reported in the previous section. Indeed, one finds that the post-jump exponents $p_r$ tend to be larger for EMC jumps than for SEC jumps; the difference $|p_{\ell}^{EMC} - 1|$ is large.
Moreover, we see that, whereas the values of $\mathcal{A}^{SEC}$ are clustered around zero, the peak of the distribution of $\mathcal{A}^{EMC}$ is clearly shifted towards negative values, as expected. Moreover, we notice that the median values of the empirical $p_r$ distribution are, respectively, $0.43$ and $0.7$ for the SEC and EMC jumps, values that are extremely close to the relaxation exponents found for the aggregated volatility profiles, and again consistent with the predictions of the Hawkes model~\ref{eq: hawkes pred} with $\theta=0.3$. 

\begin{table}
\centering
  \caption{Results of the regression of $p_\ell $, $p_r$ and $ \mathcal{A}$ against the class of each volatility time series obtained by using the news data. We use both a probit and a logit model to perform the regression.} 
  \label{tab: regressions} 
{
\def\sym#1{\ifmmode^{#1}\else\(^{#1}\)\fi}
\begin{tabular}{l*{5}{c}}
\hline\hline\\[-1.8ex] 
            &\multicolumn{1}{c}{Logit} & $\quad$ & \multicolumn{1}{c}{Probit}\\
\hline\\[-2.2ex] 
$p_{\ell}$      &      -0.432\sym{***} & $\quad$ &      -0.199\sym{***}\\
            &    (0.080)         & $\quad$ &    (0.036)    \\
[1em]            
$p_r$     &     0.469\sym{***} & $\quad$ &     0.300\sym{***} \\
            &   (0.131)         & $\quad$ &   (0.070)       \\
[1em] 

$\mathcal{A}$      &      -1.897\sym{***}  & $\quad$ &      -0.906\sym{***} \\
            &   (0.211)         & $\quad$ &    (0.104)       \\
[1em] 

const.      &      -3.623\sym{***}  & $\quad$ &      -2.001\sym{***} \\
            &   (0.110)         & $\quad$ &    (0.052)       \\
[1em]

\hline\\[-2.2ex] 
AUC      &        0.73         & $\quad$ &        0.73       \\
pseudo-$R^2_{adj}$   &       0.069         & $\quad$ &       0.070    \\
\hline\hline\\[-1.8ex] 
\multicolumn{4}{l}{\footnotesize Standard errors in parentheses. Two-tailed test.}\\
\multicolumn{4}{l}{\footnotesize \sym{***} $p<0.001$}\\
\end{tabular}
}
\end{table}

In order to confirm that exogenous and endogenous events are genuinely characterised by different volatility profiles, we use the values of $p_\ell$, $p_r$ and $\mathcal{A}$ fitted on individual profiles to perform a regression on the given \emph{a priori} EM and SE classes using both a Probit and a Logit model. The results of the two regressions are reported in Table~\ref{tab: regressions}. First of all we observe that the classification task can successfully be performed using the three selected features, as witnessed by the values of the Area Under the Curve (AUC) and of the pseudo-$R^2$. 

Moreover we see that the three features (higher pre-shock explosiveness, stronger asymmetry and faster post-shock relaxation) predicted for the EMC (marked as 1 in the regression) are indeed attested by the values of the parameters associated with each regressor.

\begin{figure}[t]
  \centering
  \includegraphics[scale=0.9]{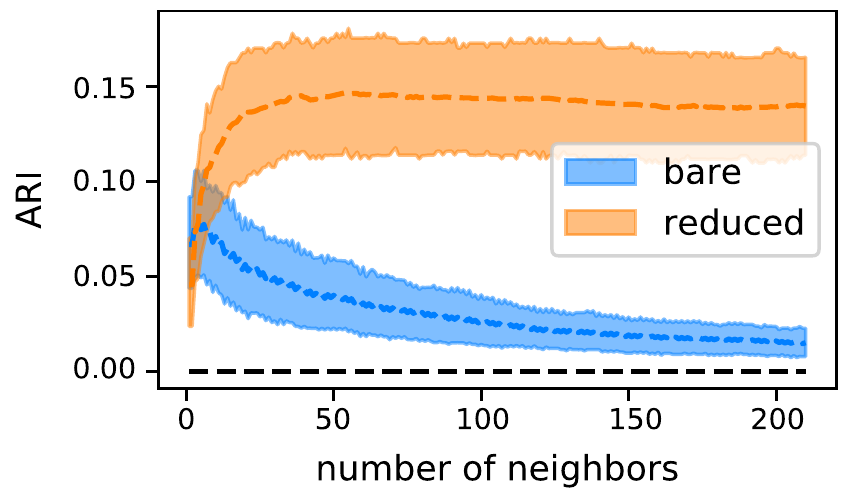} 
  \caption{Average ARI scores of a $K$-nn exploration of the space around each time series at different $K$ levels. In blue we report the results when each time series is considered in its bare form, while in red we report the results obtained when each time series is embedded in the space $(p_\ell,p_r, \mathcal{A})$. The bands are the upper and lower 0.01 quantiles of the ARI distribution for a particular $k$ when a bootstrapped subsample of the endogenous class is performed to match the number of elements in the exogenous class.}
  \label{fig: knn}
\end{figure}

In order to show that the results obtained are genuine, we randomly split our pool of jumps into a training and a test set (using a 80/20 ratio). We then train the regression models of Table~\ref{tab: regressions} using only the former set and we test in on the latter (i.e. we test the model on unseen data) by means of the AUC metric. We repeat the process 1000 times. The average values of the out-of-sample AUC coming from this experiment are  $0.72 \pm 0.03$ for both the Logit and Probit model, very close to the full sample result. 

To further corroborate that the three features we have selected provide a meaningful low dimensional embedding for the classification of jumps into EM and SE classes, we explore the neighbourhood of each volatility profile by means of a distance weighted $K$-nearest neighbors algorithm~\cite{knn_ref}. First, we perform a bootstrap down-sampling of the SEC sample in order to have the same number of endogenous and exogenous time series. For each time series, we consider, using the euclidean distance, its first $K$ neighbours and their (known) classes. We then assign each time series to the class most represented among those $K$ neighbours. We repeat the process for each time series. 

We then compare the resulting classification with the correct one. This comparison is performed using the so-called Adjusted Rand Index~\cite{ARI_ref} (ARI), which is 0 for a random classifier. In Fig.~\ref{fig: knn} we plot the result of this $K$-nn exploration at various values of $K$. We observe that, for low $K$ values, using the bare time series or its lower dimensional embedding gives comparable ARI values, which are low but both distinguishable from 0. As soon as we move away from $K=1$ (and the so called ``curse of dimensionality'' kicks in for the full time series) we see that the average ARI of the $K$-nn classification becomes 0 for the full time series, while it rapidly converges to $0.15$ for the three dimensional embedding we propose. This latter observation suggests that, in the space $(p_\ell,p_r,\mathcal{A})$, exogenous and endogenous jumps are overlapping but distinguishable clusters of points and therefore that SEC and EMC are distinguishable classes also when the intrinsic noise of the systems is not filtered out by an aggregation procedure. 

To further strengthen our results, we have also performed the very same analyses using a fitting procedure, similar to the one suggested in Refs.~\cite{sornette_books,sornette_youtube}, and we find qualitatively similar results (see the Appendix for a detailed explanation). 

\section{Conclusions}

Building upon the literature characterising the relaxation properties of financial systems after large exogenous shocks, we have argued that such fingerprints can fruitfully be used to disentangle exogenous and endogenous events, in close analogy with what has been observed in other social systems where self-exciting effects play an important role.

Using 5 years of minute by minute data collected from the Limit Order Books of 300 different NYSE stocks, we have shown that the average characteristics of clusters of abnormally large price variations in close proximity of a news release differ significantly from those occurring without any triggering event in the news feed. In particular, we have shown how the {\it average} profiles of the instantaneous volatility, past volatility and normalised past trends all display specific fingerprints that discriminate between exogenous and endogenous jumps. We have also focused on {\it individual} volatility profiles and have shown that, despite a modest signal to noise ratio, the parameters of the fitted power-laws allow one to reconstruct the classification provided by the news feed of large jumps into the ``self-excited'' and news induced class.

Whereas the existence of exogenous and endogenous types of shocks in financial systems may appear natural to many, it should be stressed that it is still a matter of intense skepticism in the current economic literature, given the difficulty to reconcile this view with the enduring Efficient Market Theory. We hope that the present work will help convince researchers that, while markets do indeed strongly and rapidly react to outstanding news, small and seemingly unimportant fluctuations may lead to a cascade of events that trigger large price jumps. In fact, most jumps appear to be of such type -- market participants do endemically interact, both directly and indirectly. The very existence of public sources of information -- such as the price itself and the Limit Order Book -- leads to global interactions and destabilising feedback loops. 

Our study can be seen as supporting a micro-structural interpretation of the excess volatility puzzle~\cite{shiller}: if large price jumps can appear out of the blue, as a result of intrinsic market fragility, then it is not surprising that prices are also too volatile. In fact, dissecting the mechanisms that lead to clusters of jumps using a multi-dimensional Quadratic Hawkes processes calibrated on tick-by-tick order data would be a very interesting follow up which we leave for future work (see \cite{fosset2021calibration}). 

Finally, while we are confident that our results have a large degree of universality, extending them to other asset classes, market places or timescales would certainly be of interest and would bolster our claim that endogenous price jumps fall in the wider class of self-excited events.  

\section*{Acknowledgments}

We thank Pierre-Philippe Cr\'epin who contributed to the early stages of this work, as well as Cecilia Aubrun, Charles-Albert Lehalle, Antoine Fosset and Iacopo Mastromatteo for fruitful discussions. We also thank Gary Kazantsev (Bloomberg), Fabrizio Lillo and Didier Sornette for encouragements, comments and suggestions.

This research was conducted within the Econophysics \& Complex Systems Research Chair, under the aegis of the Fondation du Risque, the Fondation de l’Ecole polytechnique, the Ecole polytechnique and Capital Fund Management.

\bibliography{bib}

\clearpage
\onecolumngrid
\appendix

\section{Data handling and descriptive statistics}

\subsection{Financial data preprocessing}

The full list of the stocks' tickers included in our analysis is the following: TSLA, AMZN, AAPL, MSFT, FB, NVDA, GOOGL, GOOGL, NFLX, AMD, ZM, BA, INTC, V, PYPL, ADBE, JPM, BRK/A, MA, BAC, CSCO, JNJ, DIS, UNH, QCOM, CMCSA, COST, CRM, PG, XOM, GILD, MU, T, PEP, PFE, HD, ROKU, C, AVGO, BKNG, WMT, TXN, MRNA, AMGN, CHTR, SBUX, WFC, VZ, MRK, UAL, CVX, KO, BYND, LRCX, MCD, REGN, TMUS, DOCU, ABBV, ORCL, SQ, NKE, BMY, BIIB, ATVI, PTON, AMAT, EBAY, AAL, TMO, IBM, FISV, VRTX, GS, NEE, ISRG, UBER, INTU, UTX, NOW, LLY, LULU, CRWD, UNP, ABT, TTD, HON, LMT, ILMN, MMM, MS, MELI, TWTR, LOW, AMT, MDLZ, TGT, ADP, DXCM, ADI, EQIX, GE, BLK, EA, WDAY, ADSK, DAL, MAR, CME, UPS, XLNX, CSX, DHR, CAT, SPGI, SPLK, FIS, TWLO, PM, CVS, NEM, AXP, COUP, WYNN, ORLY, TDOC, WBA, FDX, BDX, SNAP, ECL, NVAX, TJX, ETSY, PLD, F, EXPE, MCHP, SCHW, ROST, OKTA, KLAC, SWKS, ANTM, CI, DKNG, DDOG, DG, GM, MO, SPG, DLR, CMG, NKLA, ALGN, DUK, CTXS, XEL, EL, TTWO, CCI, CL, COP, OXY, HUM, EXC, CTSH, SHW, VIAC, NOC, ALXN, WORK, BSX, APD, ULTA, ZS, WDC, ENPH, MCK, D, SBAC, LUV, GPN, LYFT, ZTS, CLX, PENN, ICE, DE, W, SYK, USB, DD, KMB, MXIM, SO, DLTR, KR, DPZ, PINS, MPC, MTCH, KHC, CDNS, MNST, SNPS, AEP, LHX, INO, ZG, ITW, SEDG, NSC, PNG, TROW, FTNT, FSLY, IAC, IDXX, MET, EW, RNG, AKAM, CNC, CTAS, TIF, VLO, GD, FAST, PANW, LVGO, MMC, EOG, PAYX, WM, HCA, AZO, PSX, MCO, PGR, QRVO, TER, CSGP, FCX, TSCO, MDB, PCAR, GIS, TFC, LVS, ANSS, BAX, PSA, BK, CZR, VEEV, HPQ, VRSN, SRE, HLT, TDG, CMI, CERN, ROP, VRSK, STZ, EMR, MGM, KMI, SGEN, CPRT, TRV, DHI, DOW, CVNA, COF, SYY, PLUG, INCY, MSI, ALL, CHRW, AIG, FLT, AVB, FE, MSCI, PPG, BBY, RUN, BMRN, YUM, WMB, QDEL, ODFL, ED, LEN, PXD, PAYC, NLOK, AMTD. 

For each stock, we have a complete description of the first 6 price levels of its LOB. If we record a missing value in the volumes or the prices at any price level which is not the best bid/ask, we simply consider the associated LOB sparsity as missing and we exclude it from any calculation performed in the main text. If we record a missing value at the best bid or at the best ask, we move the last available observation forward in time. If, in doing so, we obtain an impossible price level (i.e. the price at the best bid/ask is lower/greater than the price at the second best or greater/lower than the price at the best ask/bid), we mark as missing the associated mid-price. We also mark as missing any mid-price associated with a minute when both the best bid and best ask prices are missing. We exclude from our analysis any day with more than 25 consecutive missing mid-prices. We also exclude from our analysis the days with more than 25 consecutive minutes without any recorded price movement and the days that do not have at least 300 minutes with a recorded price movement (i.e.~a return which is not 0 nor missing). To further discount the possibility of detecting a spurious jumps due to an interval of missing values, we follow Ref.~\cite{cojump_news}  and we rescale the returns computed after a missing period with the square root of the period length. For example, given the price series $p_0, p_1, -, -, p_4, p_5$, we construct the following log-returns series $\log \frac{p_1}{p_0}, \text{NA}, \text{NA},\frac{1}{\sqrt{3}} \log \frac{p_4}{p_1},\log \frac{p_5}{p_4}$.    

\subsection{Returns standardization}

\begin{figure*}
  \centering
  \includegraphics[width=0.8\linewidth]{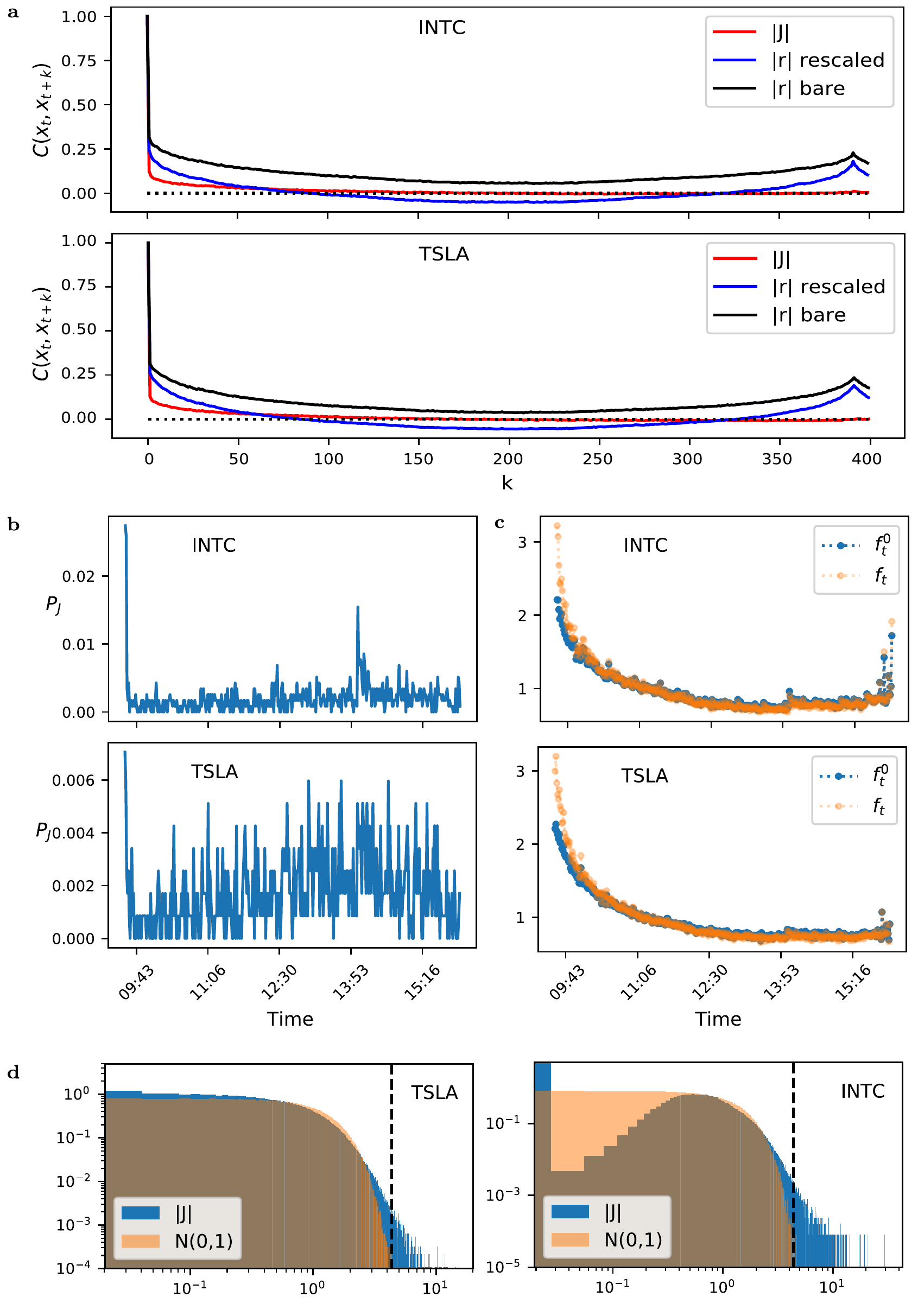}
  \caption{\textbf{(a)} Autocorrelation functions of the absolute returns $\vert r_t\vert$ (black), of the absolute rescaled returns $\vert r_t\vert/\sigma_t$ (blue) and of the absolute jump statics $\vert J_t\vert$ (red) at different lags for TSLA and INTC stocks.  \textbf{(b)} Evolution of the unconditional jump probability for TSLA and INTC stocks. \textbf{(c)} Estimated periodicity factors $f_t$ for TSLA and INTC stocks. \textbf{(d)} Probability distribution of the absolute value of the jump statics $J$ for two selected stocks. In orange we display the pdf of the absolute value of a standard normal distribution.}  
  \label{fig: appendix 1}
\end{figure*}

As mentioned in the main text, the log-returns $r_t$ time series in not suitable for the identification of price jumps and must be standardize by passing from $r_t$ to $J_t = \frac{r_t}{\sigma_t f_t}$. As a jump-robust estimator of the local volatility, we use, as suggested in Refs.~\cite{jump_test_1,jump_test_2}, the square root of the average realised bipower variation:
\[
\sigma^2_t = \frac{\pi}{2 K} \sum_{i=1}^{K} \vert r_{t-i} \vert\, \vert r_{t-i+1}\vert\; .
\]
To estimate the periodicity component $f_t$ of the volatility, we perform a two step procedure based on Refs.~\cite{jump_test_1,jump_test_2}. First of all, let's define $\hat{r}_t = \frac{r_t}{\sigma_t}$. Let $\hat{r}_{1,i}, \ldots , \hat{r}_{n_i,i}$ be the set of standardized returns having the same periodicity factor as $r_i$, i.e. the returns of a given stock, all recorded at the given time interval $i$. We define the periodicity factor $f_i$ of the time interval $i$ as:
\[
f_i = \frac{W_i}{ \sqrt{T^{-1} \sum_j W_{i-j}^2 }} \;, \quad \text{where} \quad W_i = \sqrt{1.081 \frac{\sum_j^{n_i} \Theta(-\hat{r}_{j,i}^2+x) \hat{r}_{j,i}^2}{\sum_j^{n_i} \Theta(-\hat{r}_{j,i}^2+x)}} \;.
\]
Beside being normalised so that the squared periodicity factor has mean one over any local window of length $T$, $f_i$ is a simple weighted standard deviation of the squared standardized returns $\hat{r}_{j,i}$. The weights are 1 for those squared standardize returns which are below a threshold value $x$ and 0 otherwise. To estimate the periodicity we perform a two-fold procedure. First we estimate $f^0_i$ using $x=4^2$, i.e. by excluding from the calculation those rescaled returns $\hat{r}_t$ more than 4 standard deviation away from the average. Then, using $\hat{r}_t/f_i$ (which are now very close to be normally distributed), we perform a second periodicity estimation $f^1_i$ with a threshold value $x=6.635$, i.e. the 99\% quantile of the $\chi^2$ distribution with one degree of freedom. We then define the final periodicity factor $f_t$ as $f_t = f^1_t f^0_t$. We use as periodicity cycle one day and we therefore consider as having the same periodicity factor all the returns of a stock which happen in the same minute of different days. In Panel \textbf{(c)} of Figure~\ref{fig: appendix 1} we plot the estimated periodicity factors of two different stocks. As it can be seen, the two-fold procedure we used puts a final higher factor on the first 15 minutes of the day and leaves the rest almost untouched. It is therefore less prone to detect jumps at the opening of the trading day. We also remind that, to further discount for spourious effects due to the opening and closing, we excluded from the jump pool those jumps detected in the first/last 15 minutes of the day. We also explored the possibility of having a cycle of one week (i.e. we consider as having the same periodicity factor all the returns of a stock which happen in the same minute of the same day of different weeks) but we do not find any sizable difference in the final $J_t$.

To show that the final jump statistics $J_t$ is indeed effective, in Panel \textbf{(a)} of Figure~\ref{fig: appendix 1}, we plot the autocorrelation function of $\vert J_t\vert$ for two different stocks. As it can be seen, $J_t$ does a fairly good job in taking out from the volatility most of its seasonal components as well as most of its internal dynamics. To give the reader a better understanding of the jump statistics we use, in Panel \textbf{(d)}, we plot the final probability density function of $\vert J_t\vert$ with respect to that of the absolute standard normal distribution. Finally, in Panel \textbf{(b)} we show how the unconditional jump probability $P_J$ resulting from $J_t$ evolves over time for two selected stocks.

\subsection{News data}

\begin{figure*}
  \centering
  \includegraphics[width=0.8\linewidth]{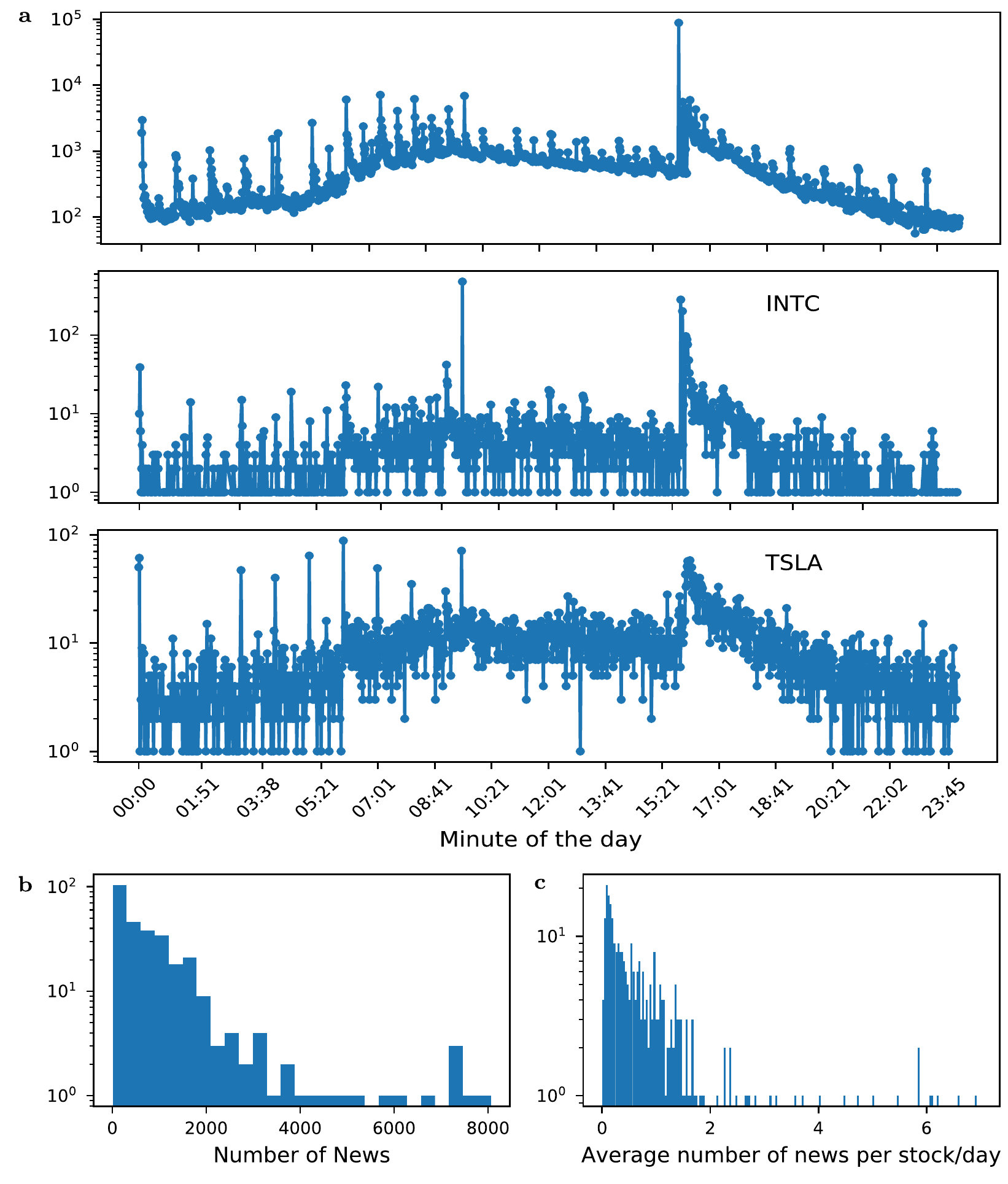}
  \caption{\textbf{(a)} Number of news recorded for each minute of the day across all stocks and for two selected stocks. \textbf{(b)} Empirical distribution of the number of news (recorded within each trading day) across stocks. \textbf{(c)} Empirical distribution of the average daily number of news across stocks.}  
  \label{fig: appendix 2}
\end{figure*}

In Panel \textbf{(a)} of Figure~\ref{fig: appendix 2} we display the number of stock-specific news for each minute of the day. In Panel \textbf{(b)} and \textbf{(c)}, we respectively report the empirical distribution of news per stock and the average number of news per stock.

\section{Power law fitting}

In this section we detail the fitting procedure used in the main text as well as the secondary fitting procedure. First of all it should be noticed that fitting power law scaling laws is a subtle topic that has been vastly debated in the literature~\cite{beran1994statistics}. Here, we are trying to fit a double power law function to very noisy data, as such every step should be done with extreme care. 

The secondary procedure we adopt is similar to the one suggested in Refs.~\cite{sornette_books,sornette_youtube}. We consider the following functional form:
\begin{equation}\label{eq: f1}
f_1(t \vert N_{\ell},N_r,p_{\ell},p_r,t_c) = \Theta(-t+t_{c}) \log \frac{N_{\ell} \;}{\vert t-t_{c} \vert^{p_{\ell}}} + \Theta(t-t_{c}) \log \frac{N_r \;}{\vert t-t_{c} \vert^{p_r}}
\end{equation}
Note that no baseline volatility $d$ is added to the fit. This is done in order to keep the problem analytically solvable. To amend for this lack of a constant parameter, instead of fitting Eq.~\ref{eq: f1} to $\log \vert J_t\vert$, we fit it against $\log \frac{\vert J_t\vert}{J_0}$, i.e. against the instantaneous volatility profile normalised for the size of the first jump of a cluster. Doing so minimizes the influence of the baseline volatility on the fit, normalises all the fits and does not modify the values of the best fitting exponents of the power laws. Performing a least square fit of Eq.~\ref{eq: f1} on an empirical volatility series $\log \frac{\vert J_{t_i}\vert}{J_0}$ means to solve the following optimization problem:
\[
\min_{N_{\ell},N_r,p_{\ell},p_r,t_c} \sum_i \left( f_1(t_i)-\frac{\vert J_{t_i}\vert}{J_0} \right)^2 \; .
\]
Calling $\vert t_i-t_{c} \vert= \Delta t_i$, $A_{\ell/r} = \log N_{\ell/r}$, $L = \sum_{i \vert t_i<t_{c}} 1$, $R = \sum_{i \vert t_i>t_{c}} 1$, $s_{\ell} = \sum_{i \vert t_i<t_{c}} \log \Delta t_i$, $s_r = \sum_{i \vert t_i>t_{c}} \log \Delta t_i$, $S_{\ell} = \sum_{i \vert t_i<t_{c}} \log^2 \Delta t_i$, $S_r = \sum_{i \vert t_i>t_{c}} \log^2 \Delta t_i$, $D_{\ell} = \sum_{i \vert t_i<t_{c}} \frac{\vert J_{t_i}\vert}{J_0}$, $D_r = \sum_{i \vert t_i<t_{c}} \frac{\vert J_{t_i}\vert}{J_0}$, setting the partial derivatives with respect to $A_{\ell/r}, p_{\ell/r}$ to zero and solving the system of equations, gives:
\[
p^\star_{\ell} = D_{\ell} \frac{L - s_{\ell}}{s^2_{\ell} - L S_{\ell}} \; , \; A^\star_{\ell} = D_{\ell} \frac{s_{\ell} - S_{\ell}}{s^2_{\ell} - L S_{\ell}} \; , \; p^\star_r = D_r \frac{R - s_r}{s^2_r - R S_r} \; , \; A^\star_r = D_r \frac{s_r - S_r}{s^2_r - R S_r} \;\; .
\]
Note that these values of the best-fitting parameters are all functions of the unknown $t^\star_c$. To find the best-fitting shock time $t^\star_c$ able to minimize the sum of the squared residuals, we perform a numerical grid search inside the open interval $(t_{\text{j}}-1, t_{\text{j}})$, where with $t_{\text{j}}$ we indicate the time of the first jump of a cluster of jumps. The lest square fit defined in this way is unique given an empirical time series $J_{t_i}$ which is also uniquely defined by an interval $[t_1,t_N]$. As such, we fix the fitting interval to a centered interval of length 160 minutes around $t_{\text{j}}$.

After finding the optimal values of $p_{\ell}$, $p_{r}$ and $\mathcal{A}$, we perform with them both the logistic regression exercise and the $k$-nn exploration detailed in the main text. As it can be seen from Table and Figure, the results we obtain are consistent with the one reported in the main text.

\begin{table}
\centering
  \caption{Results of the regression of $p^\star_\ell $, $p^\star_r$ and $ \mathcal{A}$ against the class of each volatility time series obtained by using the news data. We use both a probit and a logit model to perform the regression.} 
  \label{tab: regressions appendix} 
{
\def\sym#1{\ifmmode^{#1}\else\(^{#1}\)\fi}
\begin{tabular}{l*{5}{c}}
\hline\hline\\[-1.8ex] 
            &\multicolumn{1}{c}{Logit} & $\quad$ & \multicolumn{1}{c}{Probit}\\
\hline\\[-2.2ex] 
$p_{\ell}$      &      -15.06\sym{***} & $\quad$ &      -7.7379\sym{***}\\
            &    (2.13)         & $\quad$ &    (1.04)    \\
[1em]            
$p_r$     &     21.53\sym{***} & $\quad$ &     10.86\sym{***} \\
            &   (1.99)         & $\quad$ &   (0.97)       \\
[1em] 

$\mathcal{A}$      &      -16.91\sym{***}  & $\quad$ &      -8.24\sym{***} \\
            &   (1.73)         & $\quad$ &    (0.82)       \\
[1em] 

const.      &      -7.87\sym{***}  & $\quad$ &      -3.97\sym{***} \\
            &   (0.25)         & $\quad$ &    (0.12)       \\
[1em]

\hline\\[-2.2ex] 
AUC      &        0.82         & $\quad$ &        0.82       \\
pseudo-$R^2_{adj}$   &       0.198         & $\quad$ &       0.198    \\
\hline\hline\\[-1.8ex] 
\multicolumn{4}{l}{\footnotesize Standard errors in parentheses. Two-tailed test.}\\
\multicolumn{4}{l}{\footnotesize \sym{***} $p<0.001$}\\
\end{tabular}
}
\end{table}

\begin{figure}[t]
  \centering
  \includegraphics[width=0.45\linewidth]{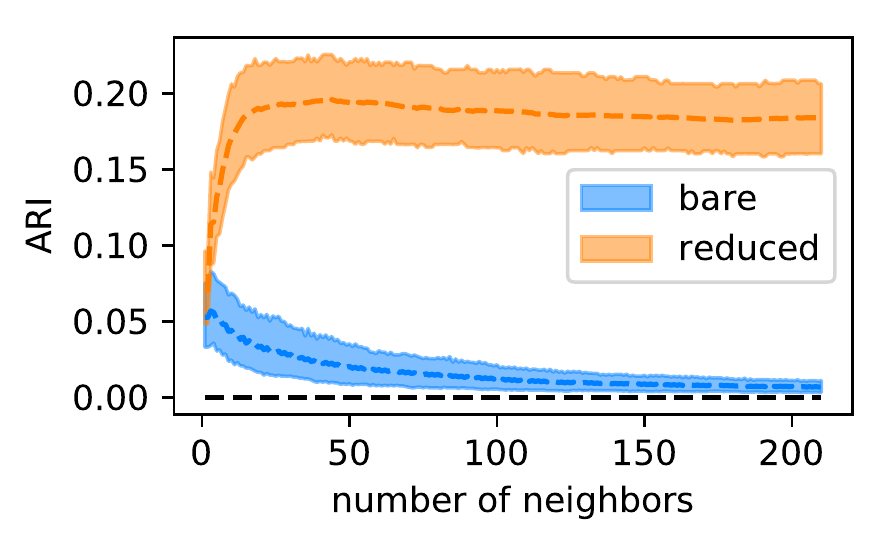} 
  \caption{Average ARI scores of a $K$-nn exploration of the space around each time series at different $K$ levels. In blue we report the results when each time series is considered in its bare form, while in red we report the results obtained when each time series is embedded in the space $(p_\ell,p_r,\vert \mathcal{A}\vert)$. The bands are the upper and lower 0.01 quantiles of the ARI distribution for a particular $k$ when a bootstrapped subsample of the endogenous class is performed to match the number of elements in the exogenous class.}
  \label{fig: knn appendix}
\end{figure}

The procedure adopted in the main text to find the best fitting parameters is a non-linear lest square performed using the SciPy Python package. Fitting directly the functional form:
\begin{equation*}
f_2(t \vert N_{\ell},N_r,p_{\ell},p_r,t_c,d ) = \frac{N_{\ell} \; \Theta(-t+t_c)}{\vert t-t_c \vert^{p_{\ell}}} + \frac{N_r \; \Theta(t-t_c)}{\vert t-t_c \vert^{p_r}} + d \; ,
\end{equation*}
on the empirical volatility series $\vert J_{t_i}\vert$ can become hard given the notoriously low signal-to-noise ratio of financial data. As such, we fit its cumulative sum $F_2(t_i) = \sum_{k=1}^i f_2(t_k)$ to $D_i = \sum_{k=1}^i \vert J_{t_k}\vert$. We restrict $t_c \in (t_{\text j}-1, t_{\text j})$ and $d>0$.

\section{Market-wide jump detection}

In order to mark a cluster of jumps as market-wide, one possibility (not used in the main text) is to compare the empirical number of clusters (of other stocks) it overlaps, against the one expected under a null-hypothesis of cluster independence. In order to create a null model of independent clusters, we perform the following randomization procedure. We call the beginning of a cluster, the time $t_{\text l}$ at which the first jump of a cluster is recorded. We call the ending of a cluster, the time $t_{L}$ at which the last jump of a cluster is recorded. Finally, we refer to $L=t_{L}-t_{\text l}$ as the length of a cluster. Given a cluster of jumps $C$, we fix its position $[t^C_{\text l},t^C_{L}]$ and we perform a numerical shuffling of all the remaining clusters beginning and ending positions so that no cluster of the same stock can overlap, the length of each cluster is preserved and no cluster may be moved outside the month/year it has been observed. Once this is done, we compare at the $0.05$ statistical significance the empirical number of cluster overlapping with $C$ against the ones expected under our null model. Whenever we observe a cluster $C$ with an higher number of overlaps then those accounted for by our null model, we mark it as market-wide.

\section{Definition of $\mathcal{A}$}

As a time asymmetry measure $\mathcal{A}$ we consider the following formula:
\begin{equation}\label{eq: time asy}
\mathcal{A} = \frac{ A_{\ell}-A_r }{A_{\ell}+A_r} \; ,    
\end{equation}
where $f^*(t)$ is the best fitting curve~\ref{eq: double pl}, $A_{\ell} = \sum_{t=t_{\text{min}}}^{t_{\text{j}}-1} f^*(t)$ and $A_r = \sum_{t=t_{\text{j}}}^{t_{\text{max}}} f^*(t)$. Equation~\ref{eq: time asy} simply compares the area of the fitted curve before the shock time $t_c^*$ with the normalised area after the shock.

\end{document}